\documentclass[12pt,onecolumn,draftclsnofoot]{IEEEtran}
\usepackage{amsfonts}
\usepackage{mathbbold}
\usepackage{mathrsfs}
\usepackage{cite}
\usepackage{graphicx,psfrag}
\graphicspath{{Figures/}}
\usepackage[cmex10]{amsmath}
\usepackage{algorithmic}
\usepackage{array}
\usepackage{mdwmath}
\usepackage{mdwtab}
\usepackage{longtable}
\usepackage{booktabs}
\usepackage{indentfirst}
\usepackage{amssymb}
\usepackage{amsmath}
\usepackage{bm}

\begin{document}

\title{Power Allocation for Cognitive Wireless Mesh Networks by Applying Multi-agent $Q$-learning Approach}

\author{\IEEEauthorblockN{Xianfu Chen$^{\dag}$$^{\ddag}$, Zhifeng
Zhao$^{\dag}$$^{\ddag}$, and Honggang Zhang$^{\dag}$$^{\ddag}$}\\

\IEEEauthorblockA{$^{\dag}$York-Zhejiang Lab for Cognitive Radio and Green Communications\\
$^{\ddag}$Department of Information Science and Electronic Engineering(ISEE)\\
Zhejiang University, Zheda Road 38, Hangzhou 310027, China\\
Email: \{chenxianfu, zhaozf, honggangzhang\}@zju.edu.cn}\\
}

\maketitle

\begin{abstract}
As the scarce spectrum resource is becoming over-crowded, cognitive
radios (CRs) indicate great flexibility to improve the spectrum
efficiency by opportunistically accessing the authorized frequency
bands. One of the critical challenges for operating such radios in a
network is how to efficiently allocate transmission powers and
frequency resource among the secondary users (SUs) while satisfying
the quality-of-service (QoS) constraints of the primary users (PUs).
In this paper, we focus on the non-cooperative power allocation
problem in cognitive wireless mesh networks (\emph{CogMesh}) formed
by a number of clusters with the consideration of energy efficiency.
Due to the SUs' selfish and spontaneous properties, the problem is
modeled as a stochastic learning process. We first extend the
single-agent $Q$-learning to a multi-user context, and then propose
a conjecture based multi-agent $Q$-learning algorithm to achieve the
optimal transmission strategies with only private and incomplete
information. An intelligent SU performs $Q$-function updates based
on the conjecture over the other SUs' stochastic behaviors. This
learning algorithm provably converges given certain restrictions
that arise during learning procedure. Simulation experiments are
used to verify the performance of our algorithm and demonstrate its
effectiveness of improving the energy efficiency.
\end{abstract}


\section{Introduction}
In wireless communications, the electromagnetic radio frequency is
the most precious resource, the use of which is regulated by
governmental agencies on a long-term basis for large geographical
regions. Currently, the frequency band is overcrowded and there
hardly exists space available for the emerging wireless services.
However, on the other hand, we are increasingly beginning to see
that the fixed spectrum allocation policy has resulted in vastly
under-utilized spectrum holes. In November 2002, the Federal
Communications Commission (FCC) published a report which shows up to
70\% of the allocated spectrum in certain measurement geographical
areas in the United States are idle in most of the time \cite{FCC}.
The limited available spectrum and the inefficiency in the spectrum
usage necessitate a new communication paradigm to exploit the
existing wireless spectrum opportunistically \cite{Ian}. New
approaches such as opportunistic spectrum access (OSA) and dynamic
spectrum access (DSA) are proposed to bridge the enormous gulf in
time and space between the regulation and the potential spectrum
efficiency. CR is a promising radio technique possessing intrinsic
capability to exploit these spectrum holes by sensing a wide range
of the frequency bands and identifying currently unused spectrum
blocks, and then communicating by an opportunistically overlaying
manner \cite{Mitola, Haykin}.

Up to now, the research on CR has already penetrated into different
types of wireless networks, and covered almost every aspect of
wireless communications \cite{Tao, Yi, Schober, Fan, Yuan}. In this
paper, we focus our emphasis on the cognitive wireless mesh
networking scenario, named as \emph{CogMesh} as described in our
previous work \cite{Tao}. One of the critical challenges in
deploying \emph{CogMesh} is how to design an efficient power
allocation scheme for the usage of detected available 'spectrum
holes' among the SUs while achieving interference-tolerable spectrum
sharing with the neighboring PUs. An efficient design is to maximize
the network performance subject to guaranteeing the PU transmissions
and the signal-to-interference-plus-noise ratio (SINR) of the SUs'
ongoing connections. The transmission power is a 'double-blade'
sword. On one hand, the higher the transmission power, the better
performance a SU can expect; on the other hand, this better
performance is obtained at the expense of not only causing higher
interference to both the PUs and the other SUs, but also increasing
power consumption. In wireless networks, the choice of transmission
power fundamentally affects the performance of multiple protocol
layers. Recently, there has been much work on formulating the power
allocation problem with cross layer design. The interested reader is
referred to \cite{Thomas} and cited references therein. But they
assume that the users are cooperative. Thus, the cross layer design
problem can be converted to the system¡¯s optimal design. In our
considered \emph{CogMesh} scenario, cooperation among the
neighboring clusters helps to quantify the tradeoff, for example if
a central entity controls the signaling in the network, it can
update and broadcast the relevant information to all clusters and
their registered SUs.

However, it's more suitable to address the power allocation of
\emph{CogMesh} within a non-cooperative game-theoretic framework,
since there are conflicting interests among the clusters.
\cite{Song} considered non-cooperative energy efficient spectrum
access for a wireless CR ad hoc network by combining an
unconstrained optimization method with a constrained partitioning
procedure. \cite{Yuan} studied the distributed multi-channel power
allocation for CR networks with strategy space to address both the
co-channel interference among SUs and the interference temperature
regulation imposed by PUs. In \cite{Fan}, Fan et al. proposed a
price based spectrum management scheme for CR networks. Assuming
that SUs repeatedly negotiating their best transmission powers and
spectrum, SUs announce prices to reflect their sensitivities to the
current interference levels, and then adjust their transmission
powers. Our work originates from this non-cooperative problem,
whereas we propose a reinforcement learning algorithm in this paper
to deal with it.

In order to formulate the non-cooperative game theoretically, we
first model the self-interest property of power allocation in
\emph{CogMesh}. Generally, the concept of reward refers to the level
of satisfaction the decision-maker receives as a return of its
performed action. We construct a reward function with the
consideration of energy-efficiency. Based on the reward function, we
model the selfish behaviors as a non-cooperative power allocation
game, that is, each SU maximizes its own reward, regardless of what
all the other SUs do. In spite of this selfish nature, it is
significant for the SUs to adapt to the environment changes since
energy efficiency is highly dependent on environmental factors like
primary users' behavior patterns and traffic QoS requirement.

Therefore, we formulate the power allocation in \emph{CogMesh} as a
stochastic learning process \cite{Junling, Richard, Yiping, Fangwen}
featured by non-cooperative game playing among the local clusters,
in which the SUs are spontaneous rational players with advanced
learning capability; but the SUs may be selfish at some extent. Then
we adopt the framework of reinforcement learning known as
$Q$-learning \cite{Richard} in this paper. As illustrated in Fig. 1,
during the learning procedure, the SU updates its strategy according
to its experience with different actions without explicit modeling
of the environment. Based on the single-agent $Q$-learning
algorithm, a multi-agent $Q$-learning is proposed to accomplish the
problem of multi-user stochastic learning. One challenge of the
proposed approach to our scenario is that the SUs do not know the
information of other SUs due to the non-cooperation among clusters.
Then the networking environment is non-stationary for all SUs and
the convergence of learning process may not be assured. To alleviate
the lack of mutual information exchange, the SUs form internal
conjectures over how the other SUs react to their present actions
with only local observations from direct interactions with the
\emph{CogMesh} environment. Learning is finished asymptotically by
appropriately making the use of past experience. Essentially, our
argument is that every rational SU has the motivation to improve its
performance even if they are selfish by nature.

Some work about reinforcement learning in CR networks have been
investigated \cite{Fu, Husheng}, where the studies are focused on
the channel allocation, which is different from the topic in this
paper. Our work is the first one toward exploring the multi-agent
$Q$-learning theory in the stochastic non-cooperative power
allocation game in CR networks, especially, \emph{CogMesh}. Compared
to the previous work, this work provides the following three key
insights:
\begin{itemize}
  \item Firstly, for the non-cooperative power allocation game in
  \emph{CogMesh}, we show that the selfish dynamics exist in the
  stochastic learning process.
  \item Secondly, we present a reinforcement learning algorithm where
  the update rule is based on SU's own private and incomplete information;
  the selfish learning dynamics converge.
  \item Thirdly, this paper also contributes to the general literature
  on multi-agent $Q$-learning. While traditional multi-agent
  $Q$-learning algorithms, such as Nash-$Q$ \cite{Junling} and CE-$Q$ \cite{Amy}
  in computer science (CS), rely on  the full information of all agents in the
  environment. This is impossible in the scenarios of wireless communication, since there
  exist conflicting interests among the users. Thereupon, we
  developed a conjecture-based multi-agent $Q$-learning.
\end{itemize}

The rest of this paper is organized as follows. In the next section,
we briefly introduce the single-agent $Q$-learning algorithm and its
extension to multi-agent scenarios. In Section III, we formulate the
non-cooperative power allocation problem as a stochastic learning
game, for which we also present the design objective and the
relevant challenging issues. In Section IV, we propose a
conjecture-based multi-agent $Q$-learning algorithm; and the
convergence of the proposed algorithm is further investigated. The
numerical results are included in Section V, verifying the validity
and efficiency of the proposed algorithm. Finally, we present in
Section VI a conclusion of this paper.

\section{Preliminaries of $Q$-Learning Algorithm}
In this section we first give a brief introduction on the
single-agent $Q$-learning algorithm, and then extend the algorithm
to multi-agent scenarios. Our description adopts standard notations
and terminologies from the framework of reinforcement learning
\cite{Richard, Junling, Eduardo}.

\subsection{Single-agent $Q$-learning}
The environment, which an agent interacts with, is typically
formulated as a finite-state Markov Decision Process (MDP). Let
$\mathcal{S}$ be a discrete set of environment states, and
$\mathcal{A}$ a discrete set of actions. At each step $t$, the agent
senses the state $s^{t}=s\in\mathcal{S}$ and selects an action
$a^{t}=a\in\mathcal{A}$ to perform. As a result, the environment
makes a transition to the new state $s^{t+1}=s'\in\mathcal{S}$
according to probability $T_{ss'}(a)$ and thereby generates a
feedback (reward) $r^{t}=r(s^{t},a)\in\textbf{R}$ passing to the
agent. This process is repeated infinitely.

The task of the agent is then to learn an optimal policy
$\pi^{*}(s)$ for each $s$, which maximizes the total expected
discounted reward over an infinite steps.
\begin{equation}
    V^{\pi}(s)=E\left[\sum\limits_{t=0}^{\infty}\beta^{t}r(s^{t},\pi(s^{t}))|s^{0}=s_{0}\right],
\end{equation}
where $s_{0}$ is the initial state, $E$ means the expectation
operator and $\beta\in[0,1)$ is the discount factor. We can rewrite
Equation (1) as \cite{Richard}
\begin{equation*}
    V^{\pi}(s)=E[r(s,\pi(s))]+\beta\sum\limits_{s'\in
    \mathcal{S}}T_{ss'}(\pi(s))V^{\pi}(s').
\end{equation*}
It has been proven that the optimal policy satisfies the Bellman
optimality equation
\begin{equation}
    V^{*}(s)=V^{\pi^{*}}(s)=\max\limits_{a\in\mathcal{A}}\left\{E[r(s,a)]+
    \beta\sum\limits_{s'\in\mathcal{S}}T_{ss'}(a)V^{*}(s')\right\}.
\end{equation}

One of the attractiveness of $Q$-learning is that it assumes no a
prior knowledge about the state transition probabilities
$T_{ss'}(a)$. We define the right-hand side of Equation (2) by,
\begin{equation}
    Q^{*}(s,a)=Q^{\pi^{*}}(s,a)=E[r(s,a)]+\beta\sum\limits_{s'\in
        \mathcal{S}}T_{ss'}(a)V^{\pi^{*}}(s').
\end{equation}
Then by Equation (2),
\begin{equation*}
    V^{*}(s)=\max\limits_{a\in\mathcal{A}}Q^{*}(s,a).
\end{equation*}
The optimal state value function $V^{*}(s)$ can be hence obtained
from $Q^{*}(s,a)$. And in turn, Equation (3) may be expressed as
\begin{equation*}
    Q^{*}(s,a)=E[r(s,a)]+\beta\sum\limits_{s'\in
        \mathcal{S}}\left\{T_{ss'}(a)\left[\max\limits_{b\in\mathcal{A}}Q^{*}(s',b)\right]\right\}.
\end{equation*}

In $Q$-learning, the agent tries to find $Q^{*}(s,a)$ in a recursive
way with the information $\langle s,a,r^{t},s'\rangle$. The updating
rule is
\begin{equation*}
    Q^{t+1}(s,a)=(1-\alpha_{t})Q^{t}(s,a)+\alpha_{t}\left[r^{t}+\beta\max\limits_{b}Q^{t}(s',b)\right],
\end{equation*}
where $\alpha_{t}\in[0,1)$ is the learning rate. Assuming that each
action is executed in each state an infinite number of times and the
learning rate $\alpha_{t}$ is decayed appropriately in a suitable
way, the $Q^{t}(s,a)$ will finally converge to $Q^{*}(s,a)$ with
probability (w.p.) 1 as $t\rightarrow\infty$ \cite{Dayan}.

\subsection{Multi-agent $Q$-learning}
Consider an $N$-agent game, each agent is equipped with a standard
$Q$-learning algorithm and learns without any cooperation with the
other agents. The received rewards and state transitions, however,
depend on the joint actions of all agents. Let $\mathcal{S}_{i}$ be
a discrete set of environment states and $\mathcal{A}_{i}$ a
discrete set of actions relevant to agent $i$. At each step $t$, the
agent senses the environment state $s_{i}^{t}=s_{i}\in
\mathcal{S}_{i}$, then independently chooses action
$a_{i}\in\mathcal{A}_{i}$. Consequently, agent $i$ receives
$r_{i}^{t}=r_{i} (s_{i}^{t},a_{1},\cdot\cdot\cdot, a_{N})$ and the
environment transits to a new state $s_{i}^{t+1}=s_{i}'\in
\mathcal{S}_{i}$ according to the fixed probabilities
$T_{s_{i}s_{i}'}(a_{1},\cdot\cdot\cdot, a_{N})$. Note that
$r_{i}^{t}$ and $T_{s_{i}s_{i}'}$ are defined over the joint actions
$(a_{1},\cdot\cdot\cdot, a_{N})$.

\section{Problem Formulation}
In this paper, we consider a non-cooperative power allocation system
in which each SU behaving as a learning agent adjusts its
transmission power level based on some reward received from the
self-interested \emph{CogMesh} environment to arrive at the optimal
strategy. The key component for describing the selfish interest is
the reward function. In this section, we first present a reward
model for the power allocation, which takes the energy-efficiency
into account. Based on the reward model, we formalize the power
allocation problem through the non-cooperative game playing.
Following that, we convert the non-cooperative playing into a
stochastic learning process. Finally, we discuss the design
objective and highlight the challenging issues.

\subsection{Reward Function and Non-cooperative Power Allocation Game}
We consider a generalized \emph{CogMesh} networking example
consisting of several specific PU links (i.e., primary transmitter
PT and primary receiver PR) and one CR network formed by a set
$\mathcal{N}=\{1,\cdot\cdot\cdot,N\}$ of SU links spatially
distributed in non-overlapping clusters (see Fig. 2). Due to
opportunistic spectrum accessing, they coexist in the same area and
share the same frequency band with bandwidth $W$ simultaneously. We
assume that each user operates in a half-duplex manner, which means
it cannot receive any signal when it's transmitting, and vice versa.
The total interference plus noise measured by any SU includes
PU-to-SU interference, SU-to-SU interference, and the Additive White
Gaussian Noise (AWGN). A SU suggests a CR link consisting of a pair
of CR nodes, and we use a SU and a CR link interchangeably.

We designate the transmission power and
Signal-to-Interference-plus-Noise Ratio (SINR) for SU $i$ by
$p_{i}(p_{i}^{\min}\leq p_{i}\leq p_{i}^{\max})$ and $\gamma_{i}$,
respectively. The other SUs' transmit power vector is denoted by
$\textbf{p}_{-i}=(p_{1},\cdot\cdot\cdot, p_{i-1},p_{i+1},\cdot
\cdot\cdot,p_{N})$. Assume that the channel gains evolve slowly with
respect to the SINR evolution, the SINR of the SU $i$ in this
problem formulation is given by
\begin{equation*}
    \gamma_{i}(p_{i},\textbf{p}_{-i})=\frac{h_{ii}p_{i}}
    {\sigma+\phi_{i}^{PU}+\sum\limits_{j\in\mathcal{N}\setminus i}h_{ji}p_{j}},
\end{equation*}
where $h_{ji}$ is the channel gain between the transmitter of SU
link $j$ and the receiver of SU link $i$, $\phi_{i}^{PU}$ denotes
the PU-to-SU interference at the receiver of SU link $i$, and
$\sigma$ is the AWGN power.

The goal of power allocation within the \emph{CogMesh} framework is
to ensure that no SU's SINR falls below its threshold
$\gamma_{i}^{*}$ chosen to guarantee adequate QoS, i.e.,
\begin{equation*}
    \gamma_{i}\geq\gamma_{i}^{*}, \forall i\in\mathcal{N}.
\end{equation*}
Furthermore, the opportunistic spectrum access enables the SUs to
transmit with overlapping spectrum and coverage with PUs, as long as
that the performance degradation induced on the PUs is tolerable. In
this paper, we consider the following power mask constraint as in
\cite{Fan}, that is, the transmission power level of SU $i$ over the
detected frequency band is constrained by
\begin{equation}
    p_{i}\leq p_{mask}, \forall i\in\mathcal{N},
\end{equation}
where $p_{mask}$ is the power mask and is given as a priori. Such a
hardware based power mask is easier to manipulate at the design
stage from a practical point of view. This is because the number of
active SUs that share the same spectrum with the PUs varies in time
and space, it is impossible to design the device to account for a
'neighbor-dependent' power mask especially in the non-cooperative
\emph{CogMesh} networking.

To implement non-cooperative power allocation in \emph{CogMesh}, one
of the most important concern is the definition of the received
reward. As mentioned above, a higher SINR at the receiver will
generally result in a lower bit error rate and hence a higher
throughput. However, achieving a high SINR requires the SU to
transmit at a high power level, which in turn causes more power
consumption as well as increases the magnitude of the interference
for other users, especially the PUs. Accordingly, we choose the
average amount of bits received correctly per unit of energy
consumption as the reward function to quantify the tradeoff (as in
\cite{Farhad}), as this brings practical and meaningful metric to
define the energy efficiency,
\begin{equation*}
    \mathcal{R}_{i}(p_{i},\textbf{p}_{-i})=
    \frac{W\log_{2}\left(1+\gamma_{i}(p_{i},\textbf{p}_{-i})/\Gamma\right)}{p_{i}}.
\end{equation*}
Here, $\Gamma$ is the gap between un-coded M-QAM and the capacity,
minus the coding gain. And we assume that CR transmitters use
variable-rate M-QAM, with a bounded probability of symbol error and
trellis coding with a nominal coding gain.

Considering the power mask constraint (4), meanwhile the maximum
transmission power level $p_{i}^{\max}$, the action set of SU $i$ is
then $\mathcal{P}_{i}=[p_{i}^{\min}, \overline{p}_{i}^{\max}]$,
where $\overline{p}_{i}^{\max}= \min(p_{i}^{\max},p_{mask})$. We
formulate the SUs' selfish behaviors with the theory of
non-cooperative game defined by a tuple $\mathcal{G}=\langle
\mathcal{N},\mathcal{P}, \{\mathcal{R}_{i}(\cdot)\}\rangle$, where
$\mathcal{P}= \mathcal{P}_{1}\times\cdot\cdot\cdot
\times\mathcal{P}_{N}$ is the action space available for all SUs.
Formally, the non-cooperative power allocation game in
\emph{CogMesh} can be defined by
\begin{equation*}
    \begin{array}{l}
        \max\limits_{p_{i}\in
        \mathcal{P}_{i}}\mathcal{R}_{i}(p_{i},\textbf{p}_{-i})\\
        \mbox{s.t. }\gamma_{i}\geq\gamma_{i}^{*},
    \end{array}
\end{equation*}
for all $i\in \mathcal {N}$. The solution of this game can be
derived in the sense of Nash Equilibrium (NE) \cite{Tirole}.

\emph{Definition 1:} A transmission power vector $(p_{1}^{*},
,\textbf{p}_{-i}^{*})$ is an NE if, for each SU $i$,
\begin{equation*}
    \mathcal{R}_{i}(p_{i}^{*},\textbf{p}_{-i}^{*})\geq\mathcal{R}_{i}(p_{i},\textbf{p}_{-i}^{*}),
    \mbox{ for all }p_{i}\in\mathcal{P}_{i}.
\end{equation*}
The following proposition shows the sufficient condition for the
existence of an NE in the game \cite{Saraydar}.

\emph{Proposition 1:} For any given $p_{mask}$ value, there is an NE
in game $\mathcal{G}$ if, for $i=1,\cdot\cdot\cdot,N$:
\begin{enumerate}
  \item The action set $\mathcal{P}_{i}$ is a closed and bounded
  convex set.
  \item The reward function $\mathcal{R}_{i}(p_{i},\textbf{p}_{-i})$ is continuous in
  $(p_{i},\textbf{p}_{-i})$ and quasi-concave in $p_{i}$.
\end{enumerate}

\subsection{Stochastic Power Allocation by Multi-agent $Q$-learning}
The wireless communication system can be considered as a
discrete-time system. In this section, we model the SUs' selfish
behaviors within stochastic game framework, in which every SU plays
the role as an intelligent agent. To be compatible with the
multi-agent $Q$-learning framework, we first discrete the continuous
action profile $\mathcal{P}_{i}=\left[
p_{i}^{\min},\overline{p}_{i}^{\max}\right]$ as the following
\begin{equation}
    p_{i}(a_{i})=\left(1-\frac{a_{i}}{m_{i}}\right)p_{i}^{\min}+
    \frac{a_{i}}{m_{i}}\overline{p}_{i}^{\max},
    a_{i}^{t}=0,\cdot\cdot\cdot,m_{i}.\nonumber\\
\end{equation}
We designate $a_{i}\in\mathcal{A}_{i}=\left\{0,\cdot\cdot\cdot,
m_{i}\right\}$ as the SU $i$'s action. Then, it's necessary to
identify the environment state, the associated reward and the next
state.
\subsubsection{State}
Since there is no cooperation among the SUs, the state should be
defined based on the local observation of the environment. At time
slot $t$, we can express the state $s_{i}^{t}$ observed by the SU
$i$ as
\begin{equation*}
    s_{i}^{t}=\left(i,\mathcal{I}_{i},p_{i}(a_{i})\right)_{t}.
\end{equation*}
Herein, $\mathcal{I}_{i}\in \{0,1\}$ specifies whether the SU $i$'s
SINR $\gamma_{i}$ at the corresponding receiver end is above or
below its threshold $\gamma_{i}^{*}$. That is,
\begin{equation*}
    \mathcal{I}_{i}=\left\{
    \begin{array}{l@{\quad}l}
      1, &\mbox{ if }\gamma_{i}\geq\gamma_{i}^{*};\\
      0, &\mbox{otherwise}.
    \end{array}
    \right.
\end{equation*}
\subsubsection{Reward}
The reward $\mathcal{R}_{i}(s_{i},a_{i},\textbf{a}_{-i})=
\mathcal{R}_{i}(a_{i},\textbf{a}_{-i})$ of SU $i$ in state $s_{i}$
is the immediate return due to the execution of action $a_{i}$ when
all the other SUs choose actions $\textbf{a}_{-i}=(a_{1},\cdot \cdot
\cdot,a_{i-1},a_{i+1}, \cdot\cdot\cdot,a_{N})$. Specifically, it is
a return of choosing power level $p_{i}(a_{i})$ in state $s_{i}$ to
ensure the transmission QoS requirement as well as to achieve the
power efficiency.
\subsubsection{Next State}
According to the definition of state $s_{i}^{t}$ defined in \emph{
1)}, we can see that the state transition from $s_{i}^{t}$ to
$s_{i}^{t+1}$ is determined by the stochastic power allocations of
all SUs.

Thus the non-cooperative game $\mathcal{G}$ is converted to the
discrete form $\mathcal{G}'=\left\langle\mathcal{N},\{{A}_{i}\},
\{\mathcal{R}_{i}\}\right\rangle$, i.e., each SU chooses the
strategy $\pi_{i}(s_{i})$ independently to maximize its total
expected discounted reward
\begin{equation*}
  \max_{\pi_{i}\in\Pi_{i}}\left\{E\left[\sum_{t=0}^{\infty}\beta^{t}
  \mathcal{R}_{i}\left(s_{i}^{t},\pi_{i}(s_{i}^{t}),\bm{\pi}_{-i}(s_{i}^{t})\right)
  \big|s_{i}^{0}=s_{i}\right]\right\},\forall i\in\mathcal{N},
\end{equation*}
where $\bm{\pi}_{-i}\left(s_{i}^{t}\right)=\left(\pi_{1}(s_{1}^{t}),
\cdot\cdot\cdot,\pi_{i-1}(s_{i-1}^{t}),\pi_{i+1}(s_{i+1}^{t}),
\cdot\cdot\cdot,\pi_{N}(s_{N}^{t})\right)$ and $\Pi_{i}$ is the set
of strategies available to SU $i$. A strategy $\pi_{i}$ of SU $i$ in
state $s_{i}$ is defined to be a probability vector $\pi_{i}(s_{i})=
[\pi_{i}(s_{i},0), \cdot\cdot\cdot,\pi_{i}(s_{i},m_{i})]$, where
$\pi_{i}(s_{i}, a_{i})$ means the probability with which the SU $i$
chooses action $a_{i}$ when in state $s_{i}$. For the case of
completely exact information about the other SUs' strategies
$\bm{\pi}_{-i}=\left(\pi_{1}, \cdot\cdot\cdot,\pi_{i-1},\pi_{i+1},
\cdot\cdot\cdot,\pi_{N}\right)$, we define the total expected
discounted reward of SU $i$ over an infinite time slots as
\begin{align*}
  &V_{i}(s_{i},\pi_{i},\bm{\pi}_{-i})\\
  &=E\left[\sum\limits_{t=0}^{\infty}\beta^{t}\mathcal{R}_{i}\left(s_{i}^{t},\pi_{i}(s_{i}^{t}),
  \bm{\pi}_{-i}(s_{i}^{t})\right)\big|s_{i}^{0}=s_{i}\right] \\
  &=E\left[\mathcal{R}_{i}\left(s_{i},\pi_{i}(s_{i}),\bm{\pi}_{-i}(s_{i})\right)\right]+\beta
  \sum\limits_{s_{i}'}T_{s_{i}s_{i}'}(\pi_{i}(s_{i}),\bm{\pi}_{-i}(s_{i}))
  V_{i}\left(s_{i}',\pi_{i},\bm{\pi}_{-i}\right),
\end{align*}
where $T_{s_{i}s_{i}'}(\cdot)$ is the state transition probability,
and
\begin{equation*}
    E\left[\mathcal{R}_{i}(s_{i},\pi_{i}(s_{i}),\bm{\pi}_{-i}(s_{i}))\right]
    =\sum_{a_{1}\in\mathcal{A}_{1}}\cdot\cdot\cdot\sum_{a_{N}\in\mathcal{A}_{N}}
    \left[\mathcal{R}_{i}(a_{i},\textbf{a}_{-i})\prod_{j=1}^{N}\pi_{j}(s_{j},a_{j})\right].
\end{equation*}

In the stochastic power allocation game, each SU behaves as an
learning agent whose task is to learn the optimal strategy
$\pi_{i}^{*}(s_{i})(i=1,\cdot\cdot\cdot,N)$ for each state $s_{i}$.
Let $\bm{\pi}_{-i}^{*}=(\pi_{1}^{*},
\cdot\cdot\cdot,\pi_{i-1}^{*},\pi_{i+1}^{*},\cdot\cdot\cdot,
\pi_{N}^{*})$.

\emph{Definition 2:} A tuple of $N$ strategies $(\pi_{i}^{*},
\bm{\pi}_{-i}^{*})$ is an NE if, for each SU $i$,
\begin{equation*}
    V_{i}(s_{i},\pi_{i}^{*},\bm{\pi}_{-i}^{*})
    \geq
    V_{i}(s_{i},\pi_{i},\bm{\pi}_{-i}^{*}),
    \mbox{ for all }\pi_{i}\in\Pi_{i}.
\end{equation*}

Every finite strategic-form game has a mixed strategy equilibrium
\cite{Tirole}, that is, there always exists an NE in our game
formulation of stochastic power allocation. The optimal strategy
satisfies the Bellman optimality equation, that is, for secondary
user $i$
\begin{align}
    &V_{i}(s_{i},\pi_{i}^{*},\bm{\pi}_{-i}^{*})\nonumber\\
    &=\max\limits_{a_{i}\in\mathcal{A}_{i}}\left\{E\left[\mathcal{R}_{i}\left(s_{i},a_{i},
    \bm{\pi}_{-i}^{*}(s_{i})\right)\right]+
    \beta\sum\limits_{s_{i}'}T_{s_{i}s_{i}'}\left(a_{i},\bm{\pi}_{-i}^{*}(s_{i})\right)
    V_{i}\left(s_{i}',\pi_{i}^{*},\bm{\pi}_{-i}^{*}\right)\right\},
\end{align}
where
\begin{align*}
    E\left[\mathcal{R}_{i}\left(s_{i},a_{i},\bm{\pi}_{-i}^{*}(s_{i})\right)\right]
    =\sum\limits_{a_{1}\in\mathcal{A}_{1}}\cdot\cdot\cdot\sum\limits_{a_{i-1}\in\mathcal{A}_{i-1}}
    \sum\limits_{a_{i+1}\in\mathcal{A}_{i+1}}\cdot\cdot\cdot\sum\limits_{a_{N}\in\mathcal{A}_{N}}
    \left[\mathcal{R}_{i}(a_{i},\textbf{a}_{-i})\prod\limits_{j=1,j\neq i}^{N}\pi_{j}^{*}(s_{j},a_{j})\right].
\end{align*}
We define the optimal $Q$-value $Q_{i}^{*}$ of SU $i$ as the current
expected reward plus its future rewards when all SUs follow the Nash
equilibrium strategies, that is,
\begin{align}
    Q_{i}^{*}(s_{i},a_{i})=E\left[\mathcal{R}_{i}\left(s_{i},
    a_{i},\bm{\pi}_{-i}^{*}(s_{i})\right)\right]+
    \beta\sum_{s_{i}'}T_{s_{i}s_{i}'}\left(a_{i},\bm{\pi}_{-i}^{*}(s_{i})\right)
    V_{i}\left(s_{i}',\pi_{i}^{*},\bm{\pi}_{-i}^{*}\right).
\end{align}
Combining equations (5) and (6), it's easy to get
\begin{align*}
    Q_{i}^{*}(s_{i},a_{i})=E\left[\mathcal{R}_{i}\left(s_{i},
    a_{i},\bm{\pi}_{-i}^{*}(s_{i})\right)\right]+
    \beta\sum_{s_{i}'}T_{s_{i}s_{i}'}\left(a_{i},\bm{\pi}_{-i}^{*}(s_{i})\right)
    \max_{b_{i}\in\mathcal{A}_{i}}Q_{i}^{*}(s_{i}',b_{i}).
\end{align*}

The multi-agent $Q$-learning process tries to find $Q_{i}^{*}(s_{i},
a_{i})$ in a recursive way using the information $\langle a_{i},
s_{i},s_{i},\pi_{i}^{t}\rangle$ $(i=1,\cdot\cdot\cdot, N)$, where
$s_{i}(=s_{i}^{t})$ and $s_{i}'(=s_{i}^{t+1})$ are the states at
time slot $t$ and $t+1$, respectively; and $a_{i}$ and $\pi_{i}^{t}$
are the SU $i$'s action taken at the end of time slot $t$ and the
transmission strategy during time slot $t$. The proposed multi-agent
$Q$-learning rule is
\begin{align}
    &Q_{i}^{t+1}(s_{i},a_{i})\nonumber\\
    &=(1-\alpha^{t})Q_{i}^{t}(s_{i},a_{i})+\alpha^{t}
    \left\{\mathcal{R}_{i}(s_{i},a_{i},\textbf{a}_{-i})\prod\limits_{j=1,j\neq i}^{N}\pi_{j}^{t}(s_{j},a_{j})+
    \beta\max_{b_{i}\in\mathcal{A}_{i}}Q_{i}^{t}(s_{i}',b_{i})\right\}.
\end{align}
where $\alpha^{t}\in[0,1)$ is the learning rate.

An intuitive explanation for Equation (7) is that, once the power
level $p_{i}(a_{i})$ is selected, the increasing quantity in the
corresponding $Q$-value is updated by combining the old value and
the new expected reward. More specifically, given the probabilities
$\{\pi_{j}^{t}(s_{j},a_{j})\}_{j=1,j\neq i}^{N}$ of the other SUs
choosing power levels $\{p_{j}(a_{j})\}_{j=1,j\neq i}^{N}$, if the
SU $i$ achieves higher reward $\mathcal{R}_{i} (a_{i},\textbf{a}
_{-i})$ when selecting power level $p_{i}(a_{i})$, then the
$Q_{i}^{t}(s_{i},a_{i})$-value is increased by a higher value.
Notice that the proposed multi-agent $Q$-learning algorithm not only
needs the SU $i$'s own information, but the strategies of the other
SUs. However, in this paper, the strategy is myopic since we assume
that there is no cooperation among the SUs.

\subsection{Design Objective and Challenging Issues}
Our aim is to design stochastic power allocation in non-cooperative
\emph{CogMesh} with self-interested SUs. The reward of each SU is a
function of the joint actions of all SUs. Accordingly, we apply the
multi-agent $Q$-learning approach to model the interaction among the
SUs' strategy decisions. Rather than choosing the best transmission
power level, a SU in the stochastic learning process chooses the
best mixed strategy. The problem is challenging due to the fact that
every SU may not be aware of the following two things during the
learning process:
\begin{enumerate}
  \item the number of SUs coexist in the system;
  \item strategies available to the other SUs.
\end{enumerate}
The SU can only observe its own information, such as the environment
state, the strategy, and the received rewards.

From Equation (7), in order to learn the optimal strategy, SU $i$
needs to know not only its own strategy, but also the other SUs'
transmission strategies $\pi_{j}^{t}(j\in \mathcal {N}\setminus i)$.
Along with the discussion, we see that the obtained multi-agent
$Q$-learning algorithm cannot solve the power allocation problem
directly because no SU can observe the competing SUs' private
information in a non-cooperative \emph{CogMesh} networking scenario.
Therefore, the challenging problem arises: \textbf{\emph{how to
design a stochastic non-cooperative power allocation scheme that
guarantees SUs learning the optimal strategies with only private and
incomplete information?}}

\section{Stochastic Power Allocation with Conjecture based Multi-agent $Q$-learning Approach}
As discussed in the previous section, the main disadvantage of the
derived multi-agent $Q$-learning algorithm is its requirement to
account for the competing SUs' strategy information. In
non-cooperative power control, however, the SUs only know what
reward they are getting from their current strategy. In this
section, we propose a stochastic non-cooperative power allocation
scheme with private and incomplete information. To make the
multi-agent $Q$-learning algorithm sensible in non-cooperative
\emph{CogMesh} networking environment, it is clear that the SU needs
to conjecture the other SUs' strategy decisions without any
coordination among the local clusters \cite{Wellman}. This motivates
the conjecture based multi-agent $Q$-learning.

\subsection{Individual Behavior and Evolution}
The goal of this paper is to design a simple non-cooperative power
allocation algorithm that requires quite limited information
exchanges among the SUs. In game-theoretic point of view, the
reached NE is based on the assumptions about what knowledge the SUs
possess and assumes that every SU's strategy will not change at the
NE. Therefore, the SUs operating at the NE can be viewed as learning
agents behaving optimally with respect to their conjectures about
the strategies of the other SUs.

From Equation (7), we can see that the SU $i$'s current expected
reward depends on both its own decision and the other SUs'
transmission policies. However, in the non-cooperative scenario, it
is hard for the SUs to obtain the information of exact transmission
strategies of their competitors. We define $c_{i}^{t}(s_{i},a_{i})
=\prod\limits_{j=1, j\neq i}^{N}\pi_{j}^{t} (s_{j},a_{j})$ for the
SU $i$ in time slot $t$, to be the conjecture representing the
aggregated effect on the $Q_{i}^{t+1}(s_{i},a_{i})$-value when all
the other SUs choosing actions $\textbf{a}_{-i}$ according to their
corresponding strategies $\bm\pi_{-i}^{t}(s_{i})=\left
(\pi_{1}^{t}(s_{1}),\cdot\cdot\cdot,\pi_{i-1}^{t}(s_{i-1}),
\pi_{i+1}^{t}(s_{i+1}),\cdot\cdot\cdot,\pi_{N}^{t}(s_{N})\right)$.
Therefore, we assume that $c_{i}^{t}(s_{i},a_{i})$ is the only
information that the SU $i$ has about the contention level of the
entire \emph{CogMesh} networking environment, because it is a metric
that the SU $i$ can easily calculate based on local observations.

Specifically, from SU $i$'s viewpoint, the probability of
experiencing environment state $s_{i}'$ is $\zeta_{i}=\pi_{i}^{t}
(s_{i},a_{i})c_{i}^{t}(s_{i},a_{i})$. In other words, the
probability that the SU $i$ receives reward $\mathcal{R}_{i}(s_{i},
a_{i},\textbf{a}_{-i})$ is $\zeta_{i}$. Let $n_{i}$ denote the
number of time slots between any two consecutive slot that SU $i$
achieves the same reward $\mathcal{R}_{i}(s_{i},a_{i},\textbf{a}
_{-i})$, then $n_{i}$ has an independent and identical distribution
(i.i.d.) with $\zeta_{i}$. Thereupon, we have $\zeta_{i}\cong
1/(1+\overline{n}_{i})$, where $\overline{n} _{i}$ is the mean value
of $n_{i}$ and can be locally computed by the SU $i$ itself through
the observation of its reward history. Since SU $i$ knows its own
transmission strategy $\pi_{i}^{t}(s_{i},a_{i})$, it can estimate
$c_{i}^{t}(s_{i},a_{i})$ through $\widetilde{c}_{i}^{t}(s_{i},a_{i})
=1/[(1+\overline{n}_{i})\pi_{i}^{t}(s_{i},a_{i})]$. Note that the
action available to SU $i$ is to choose the transmission power level
according to strategy $\pi_{i}^{t}(s_{i})$. We can express the SU
$i$'s conjecture $\widetilde{c}_{i}^{t} (s_{i},a_{i})$ as a function
of its own transmission strategy. A simple method is to deploy the
linear model, i.e.,
\begin{equation}
    \widetilde{c}_{i}^{t}(s_{i},a_{i})=\overline{c}_{i}(s_{i},a_{i})
    -\omega_{i}^{s_{i},a_{i}}\left[\pi_{i}^{t}(s_{i},a_{i})-\overline{\pi}_{i}(s_{i},a_{i})\right],
\end{equation}
where the so-called reference points \cite{Alain}, $\overline{c}_{i}
(s_{i},a_{i})$ and $\overline{\pi}_{i}(s_{i},a_{i})$, are specific
conjecture and probability, and $\omega_{i}^{s_{i},a_{i}}$ is a
positive scalar. In this paper, the reference points are considered
as exogenously given and of common knowledge. That is, SU $i$
assumes that the other SUs will observe its deviation from its
reference point $\overline{\pi} _{i}(s_{i}^{t},a_{i})$ and the
aggregate effect deviates from the reference point $\overline{c}
_{i}(s_{i},a_{i})$ by a quantity proportional to the deviation of
$\pi_{i}^{t}(s_{i},a_{i}) -\overline{\pi}_{i}(s_{i},a_{i})$.

Among different choices for capturing the impact of the competing
SUs as a function of its own strategy, the linear model shown in
Equation (8) is the simplest form one can think of. In the
following, we will show that such simple model is sufficient for the
secondary users to achieve optimal transmissions. The critical
question is how to choose the parameters $\left\{\overline{c}_{i}
(s_{i},a_{i}), \overline{\pi}_{i}(s_{i},a_{i}), \omega_{i}^{s_{i},
a_{i}}\right\}$ to achieve the optimal strategies $\pi_{i}^{*}$. We
can consider setting the parameter in Equation (8) to be:
\begin{equation*}
    \omega_{i}^{s_{i},a_{i}}=\frac{\prod\limits_{j=1,j\neq
    i}^{N}\pi_{j}^{*}(s_{j},a_{j})}{\pi_{i}^{*}(s_{i},a_{i})}.
\end{equation*}
It's very easy to verify that, if the reference points are
$\overline{c}_{i}(s_{i},a_{i})=\prod_{j=1,j\neq i}^{N}
\pi_{j}^{*}(s_{j},a_{j})$ and $\overline{\pi}_{i}(s_{i},a_{i})=
\pi_{i}^{*}(s_{i},a_{i})$, we have $\widetilde{c}_{i}^{*}(s_{i},
a_{i})=\prod_{j=1,j\neq i}^{N} \pi_{j}^{*}(s_{j},a_{j})$. Therefore,
such configuration of the conjectures $\widetilde{c}_{i} ^{*}$ and
the strategies $\pi_{i}^{*}$ achieve the optimal transmission. In
non-cooperative learning scenarios, SUs learn when they modify their
conjectures based on the new observations. Specifically, we first
allow the SUs to revise their reference points based on their past
local observations. We propose a simple rule for the SUs to update
their reference points. In time slot $t$, the SU $i$ set
$\overline{c}_{i}(s_{i},a_{i})$ and $\overline{\pi}_{i}
(s_{i},a_{i})$ to be $c_{i}^{t-1}(s_{i},a_{i})$ and
$\pi_{i}^{t-1}(s_{i},a_{i})$. That is, Equation (8) becomes
\begin{equation}
    \widetilde{c}_{i}^{t}(s_{i},a_{i})=c_{i}^{t-1}(s_{i},a_{i})-
    \omega_{i}^{s_{i},a_{i}}\left[\pi_{i}^{t}(s_{i},a_{i})-\pi_{i}^{t-1}(s_{i},a_{i})\right],
\end{equation}
for $i\in\mathcal{N}$.

\subsection{Conjecture based $Q$-value Updating}
Eventually, the multi-agent $Q$-learning updating rule in Equation
(7) is modified as following,
\begin{equation}
    Q_{i}^{t+1}(s_{i},a_{i})=(1-\alpha^{t})Q_{i}^{t}(s_{i},a_{i})+\alpha^{t}
    \left\{\widetilde{c}_{i}^{t}(s_{i},a_{i})\mathcal{R}_{i}(s_{i},a_{i},\textbf{a}_{-i})+
    \beta\max_{b_{i}\in\mathcal{A}_{i}}Q_{i}^{t}\left(s_{i}',b_{i}\right)\right\}.
\end{equation}
The SU $i$ updates its $Q$-values only with its own information
using Equation (10) during the stochastic learning process. To avoid
observing the other SUs' private strategy information, the SU $i$
conjectures about how its competitors' strategy decisions vary in
response to its own actions.

The purpose of stochastic power allocation is to improve performance
by explicitly balancing two competing objectives: 1) searching for
better transmission power level (exploration) and 2) gathering as
much reward as possible (exploitation), such that the SU not only
reinforces the evaluation of the power level it already knows to be
good but also explores new one. Though $\epsilon$-greedy selection
\cite{Eduardo} is an efficient method of balancing exploration and
exploitation in reinforcement learning. One drawback is that it
chooses equally among all available actions when it explores. This
implies that the worst action is as likely to be chosen as the best
one.

An alternative solution is to vary the action probabilities as a
graded function of the $Q$-value. The greedy action is given the
highest selection probability, but all the others are ranked and
weighted according to their $Q$-values. The most common method is to
use a Boltzmann distribution. The SU $i$ chooses action $a_{i}$ in
state $s_{i}$ at time step $t$ with probability \cite{Richard},
\begin{equation}
    \pi_{i}^{t}(s_{i},a_{i})=\frac{e^{Q_{i}^{t}(s_{i},a_{i})/\tau}}
    {\sum\limits_{b\in \mathcal{A}_{i}}e^{Q_{i}^{t}(s_{i},b)/\tau}},
\end{equation}
where $\tau$ is a positive parameter called the temperature. High
temperatures cause the action probabilities to be all nearly equal.
Low temperatures cause big difference in selection probabilities for
actions differ in their $Q$-values.

Now, the steps concerning power allocation corresponding to the
conjecture-based multi-agent $Q$-learning algorithm are summarized as follows:\\
\begin{longtable}{p{0.98\textwidth}}\toprule
    \begin{center}\textbf{Algorithm:} Conjecture based Multi-agent $Q$-learning
    Algorithm for SU $i$\end{center}\\
    \midrule
    \endfirsthead
    \endhead
    \endfoot
        \textbf{Initialization:}\\
            \begin{quote}Let $\emph{t}=0$,\end{quote}\\
            \begin{quote}\textbf{For} each $s_{i}$, $a_{i}$ \textbf{Do}\end{quote}\\
                \begin{quote}\begin{quote}
                Initialize strategy $\pi_{i}^{t}(s_{i},a_{i})$, $Q$-values $Q_{i}^{t}(s_{i},a_{i})$,
                and the parameter $\omega_{i}^{s_{i},a_{i}}>0$.
                \end{quote}\end{quote}\\
            \begin{quote}\textbf{End For}\end{quote}\\
        Evaluate the initial state $s_{i}=s_{i}^{t}$.\\

        \textbf{Learning:}\\
            \begin{quote}\textbf{Loop}\end{quote}\\
            \begin{quote}
            \begin{enumerate}
            \renewcommand{\labelenumi}{(\theenumi)}
                    \item Choose action $a_{i}$ according to $\pi_{i}^{t}(s_{i})$.
                    \item Measure the SINR $\gamma_{i}$ with the feedback information of the intended secondary receiver.
                    Construct the current environment state $s_{i}'=s_{i}^{t+1}$ by identifying the transmission power level,
                    and comparing $\gamma_{i}$ with the threshold $\gamma_{i}^{*}$.
                    \item If $\gamma_{i}\geq\gamma_{i}^{*}$, then a reward $\mathcal{R}_{i}(s_{i},a_{i},\textbf{a}_{-i})$
                    can be achieved; otherwise, the receiver can not receive correctly, thus obtains zero reward.
                    \item Update $Q_{i}^{t+1}(s_{i},a_{i})$ based on $\widetilde{c}_{i}^{t}(s_{i},a_{i})$ according
                    to $Q_{i}^{t+1}(s_{i},a_{i})=\left(1-\alpha^{t}\right)Q_{i}^{t}(s_{i},a_{i})+\alpha^{t}\big\{
                    \widetilde{c}_{i}^{t}(s_{i},a_{i})\mathcal{R}_{i}(s_{i},a_{i},\textbf{a}_{-i})+
                    \beta\max\limits_{b_{i}\in\mathcal{A}_{i}}Q_{i}^{t}(s_{i}',b_{i})\big\}$.
                    \item Update the strategy $\pi_{i}^{t+1}(s_{i},a_{i})=e^{Q_{i}^{t+1}(s_{i},a_{i})/\tau}\big/
                    \sum\limits_{b_{i}\in\mathcal{A}_{i}}e^{Q_{i}^{t+1}(s_{i},b_{i})/\tau},\mbox{ for all }a_{i}\in\mathcal{A}_{i}$.
                    \item Update the conjecture $\widetilde{c}_{i}^{t+1}(s_{i},a_{i})=c_{i}^{t}(s_{i},a_{i})-
                    \omega_{i}^{s_{i},a_{i}}\left[\pi_{i}^{t+1}(s_{i},a_{i})-\pi_{i}^{t}(s_{i},a_{i})\right]$.
                    \item $s_{i}=s_{i}^{t+1}$.
            \end{enumerate}
            \end{quote}\\

        \begin{quote}\textbf{End Loop}\end{quote}\\
    \bottomrule
\end{longtable}

Next, we are interested in the convergence of this algorithm. Our
proof relies on the following lemma by Szepesvari and Littman
\cite{Szepesvari}, which establishes the convergence of a general
$Q$-learning process updated by a pseudo-contraction operator. Let
$\bm{\mathcal{Q}}$ be the space of all $Q$-values.

\emph{Lemma:} Assume that $\alpha^{t}$ in Equation (10) satisfies
the sufficient conditions of Theorem in \cite{Dayan}, and the
mapping $\mathcal{H}^{t}:\bm{\mathcal{Q}}\rightarrow\bm{\mathcal
{Q}}$ meets the following condition: there exists a number $0<
\lambda<1$ and a sequence $\xi^{t}\geq 0$ converging to zero w.p. 1,
such that $\parallel\mathcal{H}^{t}Q^{t}-\mathcal{H}^{t}Q^{*}
\parallel\leq\lambda\parallel Q^{t}-Q^{*}\parallel+\xi^{t}$ for all
$Q^{t}\in\bm{\mathcal{Q}}$ and $Q^{*}=E\left[\mathcal{H}^{t}Q^{*}
\right]$, then the iteration defined by
\begin{equation*}
    Q^{t+1}=(1-\alpha^{t})Q^{t}+\alpha^{t}(\mathcal{H}^{t}Q^{t}),
\end{equation*}
converges to $Q^{*}$ w.p. 1.

For an $N$-player stochastic game, we define the operator $\mathcal
{H}^{t}$ as follows.

\emph{Definition 3:} Let $Q^{t}=(Q_{1}^{t},\cdot\cdot\cdot,Q_{N}
^{t})$, where $Q_{i}^{t}\in\bm{\mathcal{Q}}_{i}$ for $i=1,\cdot\cdot
\cdot,N$, and $\bm{\mathcal{Q}}=\bm{\mathcal{Q}}_{1}\times\cdot\cdot
\cdot\times\bm{\mathcal{Q}}_{N}$. $\mathcal{H}^{t}:\bm{\mathcal{Q}}
\rightarrow\bm{\mathcal{Q}}$ is a mapping on the complete metric
space $\bm{\mathcal{Q}}$ into $\bm{\mathcal{Q}}$, $\mathcal{H}^{t}
Q^{t}=(\mathcal{H}^{t}Q_{1}^{t},\cdot\cdot\cdot,\mathcal{H}^{t}
Q_{N}^{t})$, where
\begin{equation*}
    \mathcal{H}^{t}Q_{i}^{t}=\widetilde{c}_{i}^{t}(s_{i},a_{i})
    \mathcal{R}_{i}(s_{i},a_{i},\textbf{a}_{-i})+\beta\max_{b_{i}
    \in\mathcal{A}_{i}}Q_{i}^{t}\left(s_{i}',b_{i}\right).
\end{equation*}
Then we proceed to prove that $Q^{*}=E[\mathcal{H}^{t}Q^{*}]$.

\emph{Proposition 2:} For an $N$-player stochastic game, $Q^{*}=
E[\mathcal {H}^{t}Q^{*}]$, where $Q^{*}=(Q_{1}^{*},\cdot\cdot\cdot,
Q_{N} ^{*})$.

\emph{Proof:} Since
\begin{align*}
    Q_{i}^{*}(s_{i},a_{i})&=E\left[\mathcal{R}_{i}\left(s_{i},
    a_{i},\bm{\pi}_{-i}^{*}(s_{i})\right)\right]+
    \beta\sum_{s_{i}'}T_{s_{i}s_{i}'}\left(a_{i},\bm{\pi}_{-i}^{*}(s_{i})\right)
    \max_{b_{i}\in\mathcal{A}_{i}}Q_{i}^{*}\left(s_{i}',b_{i}\right)\\
    &=\sum_{s_{i}'}T_{s_{i}s_{i}'}\left(a_{i},\bm{\pi}_{-i}^{*}(s_{i})\right)
    \left\{\prod\limits_{j=1,j\neq i}^{N}\pi_{j}^{*}(s_{j},a_{j})\mathcal{R}_{i}(s_{i},a_{i},\textbf{a}_{-i})+
    \beta\max_{b_{i}\in\mathcal{A}_{i}}Q_{i}^{*}\left(s_{i}',b_{i}\right)\right\}.
\end{align*}
From Equation (9), $\widetilde{c}_{i}^{*}(s_{i},a_{i})=\prod
\limits_{j=1,j\neq i}^{N}\pi_{j}^{*}(s_{j},a_{j})$. Thus,
\begin{align*}
    Q_{i}^{*}(s_{i},a_{i})=E\left[\mathcal {H}^{t}Q^{*}(s_{i},a_{i})\right],
\end{align*}
for all $s_{i}$ and $a_{i}$. $\blacksquare$

We further define the distance between two $Q$-values.

\emph{Definition 4:} For any $Q, Q'\in\bm{\mathcal {Q}}$, we define
\begin{equation*}
    \left\| Q-Q'\right\|\doteq\max\limits_{i}
    \max\limits_{s_{i}}\max\limits_{a_{i}}\left|
    Q_{i}(s_{i},a_{i})-Q_{i}'(s_{i},a_{i})\right|.
\end{equation*}

\emph{Proposition 3:} $\mathcal{H}^{t}$ is a contraction mapping
operator.

\emph{Proof:} The proof is given in Appendix.

We can now present our main result in this paper that the learning
process induced by \textbf{Algorithm} converges.

\emph{Theorem:} Regardless of any initial value chosen for
$Q_{i}^{0}(s_{i},a_{i})$, if $\tau$ is sufficiently large,
\textbf{Algorithm} converges.

\emph{Proof:} The proof is the direct application of \emph{Lemma},
which establishes the convergence given two conditions. First,
$\mathcal{H}^{t}$ is a contraction mapping operator, by
\emph{Proposition 3}. Second, the fixed point condition, $Q^{*}=
E[\mathcal {H}^{t}Q^{*}]$, is ensured by \emph{Proposition 2}.
Therefore, the learning process expressed by Equation (10)
converges.

\section{Numerical Results}
To demonstrate the performance of the proposed conjecture based
multi-agent $Q$-learning algorithm, we present simulation
experiments of a hybrid \emph{CogMesh} consisting of one PU network
and one CR network. Users in \emph{CogMesh} are uniformly
distributed over a $300\mbox{m}\times300\mbox{m}$ square area, and
share the same frequency band with bandwidth of $W=1\mbox{MHz}$. The
links can communication directly if the distance between transmitter
and the corresponding receiver is no more than $30$m. The time is
divided into slots, each of length $10$ms. During each time slot,
each PU attempts to transmit with a probability of $\kappa$, the
PU's behavior factor. It's supposed that the PUs have only one
transmission power level of $200\mbox{mW}$, the AWGN power
$\sigma=10^{-7}\mbox{mW}$, and $\Gamma=1$. Also, we set the power
mask to be 200$mW$ for all SUs. The link gains used in this paper
are given by
\begin{eqnarray*}
    h=KF\left(\frac{d}{d_{0}}\right)^{-n}, \mbox{ for } d>d_{0},
\end{eqnarray*}
where $K$ is a constant set to be $10^{-6}$, the shadowing factor
$F$ is a random number and is independent and identically generated
from a lognormal distribution with a mean of $0\mbox{dB}$ and
variance $6\mbox{dB}$, $d$ is the physical distance between
transmitter and receiver, $d_{0}$ is the reference distance, and $n$
is the path loss exponent. In the whole simulation process, we set
$d_{0}=1$ and $n=4$. And we here point out that all simulated curves
in this paper show the average over $200$ episodes.

As for the proposed conjecture based multi-agent $Q$-learning
algorithm, it's implemented by each SU with a discount factor
$\beta=0.9$. And we use the following learning rate
\begin{equation*}
    \alpha^{t}=\frac{\alpha^{0}}{\theta^{t}},
\end{equation*}
where $\alpha^{0}\in[0,1)$ is the initial learning rate, and
$\theta>1$ is a scalar. Like any other learning scheme, the SUs need
a learning phase to learn the optimal transmission strategies under
the assumption that each SU can perfectly conjecture the probability
$\prod\limits_{j=1, j\neq i}^{N}\pi_{j}^{t} (s_{j},a_{j})$ during
each time slot. However, once the strategies are acquired, the SUs
take only one iteration to reach the optimal energy-efficient
transmission configuration, when starting at any initial environment
states $s_{i}(i=1,\cdot \cdot\cdot,N)$. The major concern for our
proposed algorithm is the convergence speed of the stochastic
learning dynamics. We first simulate a relatively simple networking
scenario consisting of three pairs of SU links coexisting with three
pairs of PU links with a behavior factor $\kappa=0.5$. The SUs have
two transmission power levels $\{100\mbox{mW}, 200\mbox{mW}\}$. That
is, in the proposed algorithm, $m_{i}=1$ and $\mathcal{N}= \{1,2\}$.

Without the loss of generality, we take SU $1$ for example. Fig. 3
and Fig. 4 show the simulation results for different $\alpha^{0}$
and $\tau$, which indicate that the proposed algorithm converges. We
can also see from the Fig. 3 that larger $\tau$ results in worse
expected reward. This is because exploration lasts for a longer time
even if the best power level achieving optimal transmission was
already visited. Thus, during the learning process, the SU should
set a sufficiently large temperature to balance the tradeoff between
exploration and exploitation or has to dynamically adjust it. The
curves in Fig. 4 illustrate that when $\tau$ is small, for smaller
$\alpha^{0}$ the convergence performance is worse. Since the
$Q$-values converges slowly, then still exploration phases dominates
the learning procedure, which may lead to decreasing the
opportunities of achieving optimal transmission configuration on
average. Overall, the performance of our proposed algorithm is good
when choosing a suitable learning rate $\alpha^{0}$. If the
algorithm is deployed by the SUs in \emph{CogMesh} environment,
$\alpha^{0}$ has to be chosen in advance.

Next, for a more general case, we consider that the CR network
consists of six SUs co-locating with five PUs. The PUs attempt to
transmit with a probability $\kappa=1$. Each SU has multiple
transmission power levels. The discrete transmission power levels
the SUs used are in the range from $100\mbox{mW}$ to $200\mbox{mW}$
equally spaced by $20\mbox{mW}$. We compare the expected rewards of
SUs achieved by the proposed algorithm with the system's optimum
$\mathcal{R}_{i}^{opt}=\max\limits_{\textbf{p}}\mathcal{R}_{i}
(\textbf{p})$ in Fig. 5. It can be seen from the graph that the
achieved performance is close to the optimum and the performance
loss is no more than $25\%$ on the average.

Fig. 6 depicts the expected rewards of the six secondary users
versus the PU's behavior factor $\kappa$ under the same networking
environment assumptions as in Fig. 5. As expected, a higher $\kappa$
results in higher interference caused by the PUs to the SUs, i.e.,
the expected rewards are degraded.

\section{Conclusion}
In this paper, we have studied the non-cooperative power allocation
problem specifically in \emph{CogMesh} which is modeled as a
stochastic learning process. We extend the single-agent $Q$-learning
algorithm to a multi-user context. Due to the non-cooperation among
the local clusters, a conjecture based multi-agent $Q$-learning
approach is proposed to reach the optimal transmission strategies
with only private and incomplete information. The learning SU
performs $Q$-function updating based on the conjecture about other
SUs' behaviors over the current $Q$-values. This learning algorithm
provably converges given certain restrictions that arise during
learning procedure, and the simulations demonstrate the
effectiveness of the algorithm to improve energy efficiency. The
results in this paper provide us with a new approach to design the
protocols for the non-cooperative CR networks.

\section*{Acknowledgment}
The authors would like to thank the anonymous reviewers for their
valuable comments and suggestions, which have helped improve the
quality of this paper.

\setcounter{equation}{0}
\renewcommand\theequation{A-\arabic{equation}}
\appendix
\section{Appendix}

Proof of Proposition 3.

\emph{Proof:}
\begin{align*}
    &\left\|\mathcal{H}^{t}Q-\mathcal{H}^{t}Q'\right\|\\
    &=\max\limits_{i}\max\limits_{s_{i}}\max\limits_{a_{i}}
    \left|\mathcal{H}^{t}Q_{i}(s_{i},a_{i})-\mathcal{H}^{t}Q_{i}'(s_{i},a_{i})\right|\\
    &=\max\limits_{i}\max\limits_{s_{i}}\max\limits_{a_{i}}
    \big|\left[\widetilde{c}_{i}(s_{i},a_{i})-\widetilde{c}_{i}'(s_{i},a_{i})\right]
    \mathcal{R}_{i}(s_{i},a_{i},\textbf{a}_{-i})+\\
    &\qquad\qquad\qquad\qquad\beta\left[\max_{b_{i}\in\mathcal{A}_{i}}Q_{i}\left(s_{i}',b_{i}\right)-
    \max_{b_{i}\in\mathcal{A}_{i}}Q_{i}'\left(s_{i}',b_{i}\right)\right]\big|\\
    &\leq\max\limits_{i}\max\limits_{s_{i}}\max\limits_{a_{i}}
    \big|\left[\widetilde{c}_{i}(s_{i},a_{i})-\widetilde{c}_{i}'(s_{i},a_{i})\right]
    \mathcal{R}_{i}(s_{i},a_{i},\textbf{a}_{-i})\big|+\\
    &\quad\max\limits_{i}\max\limits_{s_{i}}
    \beta\big|\max_{b_{i}\in\mathcal{A}_{i}}Q_{i}\left(s_{i}',b_{i}\right)-
    \max_{b_{i}\in\mathcal{A}_{i}}Q_{i}'\left(s_{i}',b_{i}\right)\big|\\
    &\leq\max\limits_{i}\max\limits_{s_{i}}\max\limits_{a_{i}}
    \left|\left[\widetilde{c}_{i}(s_{i},a_{i})-\widetilde{c}_{i}'(s_{i},a_{i})\right]
    \mathcal{R}_{i}(s_{i},a_{i},\textbf{a}_{-i})\right|+\beta\left\|Q-Q'\right\|.
\end{align*}
We discuss the first item $\left[\widetilde{c}_{i}(s_{i},a_{i})-
\widetilde{c}_{i}'(s_{i},a_{i})\right]\mathcal{R}_{i}(s_{i},a_{i},
\textbf{a}_{-i})$ in the last inequality above. Due to the fact that
the reference points are exogenously given and of common knowledge,
then we have
\begin{align}
    \left[\widetilde{c}_{i}(s_{i},a_{i})-\widetilde{c}_{i}'(s_{i},a_{i})\right]
    \mathcal{R}_{i}(s_{i},a_{i},\textbf{a}_{-i})
    =-\omega_{i}^{s_{i},a_{i}}\left[\pi_{i}(s_{i},a_{i})-
    \pi_{i}'(s_{i},a_{i})\right]\mathcal{R}_{i}(s_{i},a_{i},\textbf{a}_{-i})
    \label{A.1}
\end{align}

We first concentrate on the item $\pi_{i}(s_{i},a_{i})$ in Equation
(A-1). By applying Equation (11), we have
\begin{equation*}
    \pi_{i}(s_{i},a_{i})=\frac{e^{Q_{i}(s_{i},a_{i})/\tau}}
    {\sum\limits_{b\in \mathcal{A}_{i}}e^{Q_{i}(s_{i},b)/\tau}}.
\end{equation*}
When $\tau$ is sufficiently large, we get
\begin{equation*}
    e^{Q_{i}(s_{i},a_{i})/\tau}=1+\frac{Q_{i}(s_{i},a_{i})}{\tau}+
    \vartheta\left(\frac{Q_{i}(s_{i},a_{i})}{\tau}\right),
\end{equation*}
where $\vartheta\left(\frac{Q_{i}(s_{i},a_{i})}{\tau}\right)$ is a
polynomial of order $\mathcal {O}\left(\left(\frac{Q_{i}
(s_{i},a_{i})}{\tau}\right)^{2}\right)$. It's very easy to verify
that
\begin{align}
    \pi_{i}(s_{i},a_{i})=\frac{1}{m_{i}+1}
    +\frac{Q_{i}(s_{i},a_{i})}{(m_{i}+1)\tau}
    +\varrho\left(\{Q_{i}(s_{i},b)\}_{b}\right),
\end{align}
where $\varrho\left(\{Q_{i}(s_{i},b)\}_{b}\right)$ is the polynomial
of smaller order than $\mathcal {O}\left(\frac{Q_{i}(s_{i},a_{i})}
{\tau}\right)$. Note that the coefficient of the polynomial is
independent of the $Q$-value. Similarly,
\begin{align}
    \pi_{i}'(s_{i},a_{i})=\frac{1}{m_{i}+1}
    +\frac{Q_{i}'(s_{i},a_{i})}{(m_{i}+1)\tau}
    +\varrho\left(\{Q_{i}'(s_{i},b)\}_{b}\right).
\end{align}

Substituting Equations (A-2) and (A-3) to Equation (A-1) establishes
\begin{align*}
    &\left[\widetilde{c}_{i}(s_{i},a_{i})-\widetilde{c}_{i}'(s_{i},a_{i})\right]
    \mathcal{R}_{i}(s_{i},a_{i},\textbf{a}_{-i})\\
    &=-\omega_{i}^{s_{i},a_{i}}\mathcal{R}_{i}(s_{i},a_{i},\textbf{a}_{-i})
    \left[\frac{Q_{i}(s_{i},a_{i})}{(m_{i}+1)\tau}-\frac{Q_{i}'(s_{i},a_{i})}{(m_{i}+1)\tau}+
    \varrho\left(\{Q_{i}(s_{i},b)\}_{b}\right)-\varrho\left(\{Q_{i}'(s_{i},b)\}_{b}\right)\right]\\
    &=-\mathcal{C}_{i}(s_{i},a_{i})\left\{\frac{1}{m_{i}+1}\left[\frac{Q_{i}(s_{i},a_{i})}{\tau}-
    \frac{Q_{i}'(s_{i},a_{i})}{\tau}\right]+\varrho\left(\{Q_{i}(s_{i},b)\}_{b}\right)-
    \varrho\left(\{Q_{i}'(s_{i},b)\}_{b}\right)\right\}.
\end{align*}
That is, we can always take a sufficiently large $\tau$ such that
\begin{align*}
    \left|\left[\widetilde{c}_{i}(s_{i},a_{i})-\widetilde{c}_{i}'(s_{i},a_{i})\right]
    \mathcal{R}_{i}(s_{i},a_{i},\textbf{a}_{-i})\right|
    \leq\frac{1-m_{i}\beta}{m_{i}+1}\left|\left[Q_{i}(s_{i},a_{i})-Q_{i}'(s_{i},a_{i})\right]\right|,
\end{align*}
which implies
\begin{align*}
    \left\|\mathcal{H}^{t}Q-\mathcal{H}^{t}Q'\right\|
    &\leq\max\limits_{i}\max\limits_{s_{i}}\max\limits_{a_{i}}
    \frac{\beta}{m_{i}+1}\left|\left[Q_{i}(s_{i},a_{i})-Q_{i}'(s_{i},a_{i})\right]\right|
    +\beta\left\|Q-Q'\right\|\\
    &=\frac{1+\beta}{m_{i}+1}\left\|Q-Q'\right\|.
\end{align*}
Therefore, $\mathcal{H}^{t}$ is a contraction mapping operator. This
concludes the proof. $\blacksquare$

\bibliographystyle{IEEEtranS}
\bibliography{IEEEabrv,reference}

\newpage

\begin{figure}
  \centering
  \includegraphics[width=0.7\textwidth]{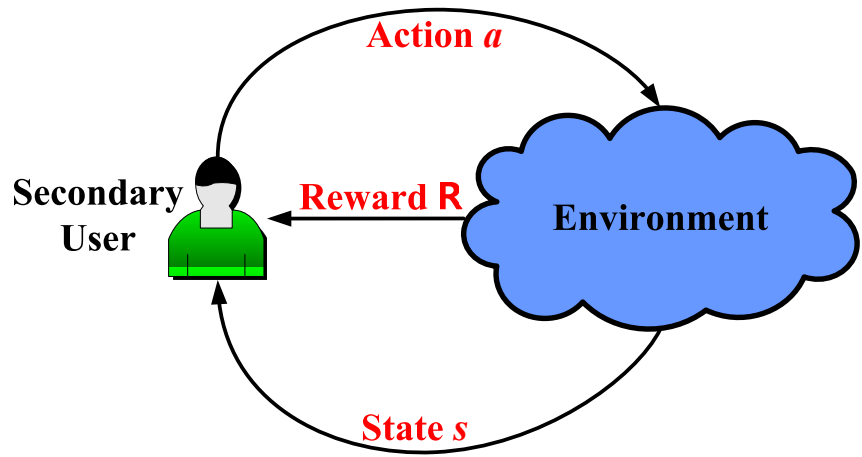}
  \caption{Reinforcement learning.}
\end{figure}

\begin{figure}
  \centering
  \includegraphics[width=0.7\textwidth]{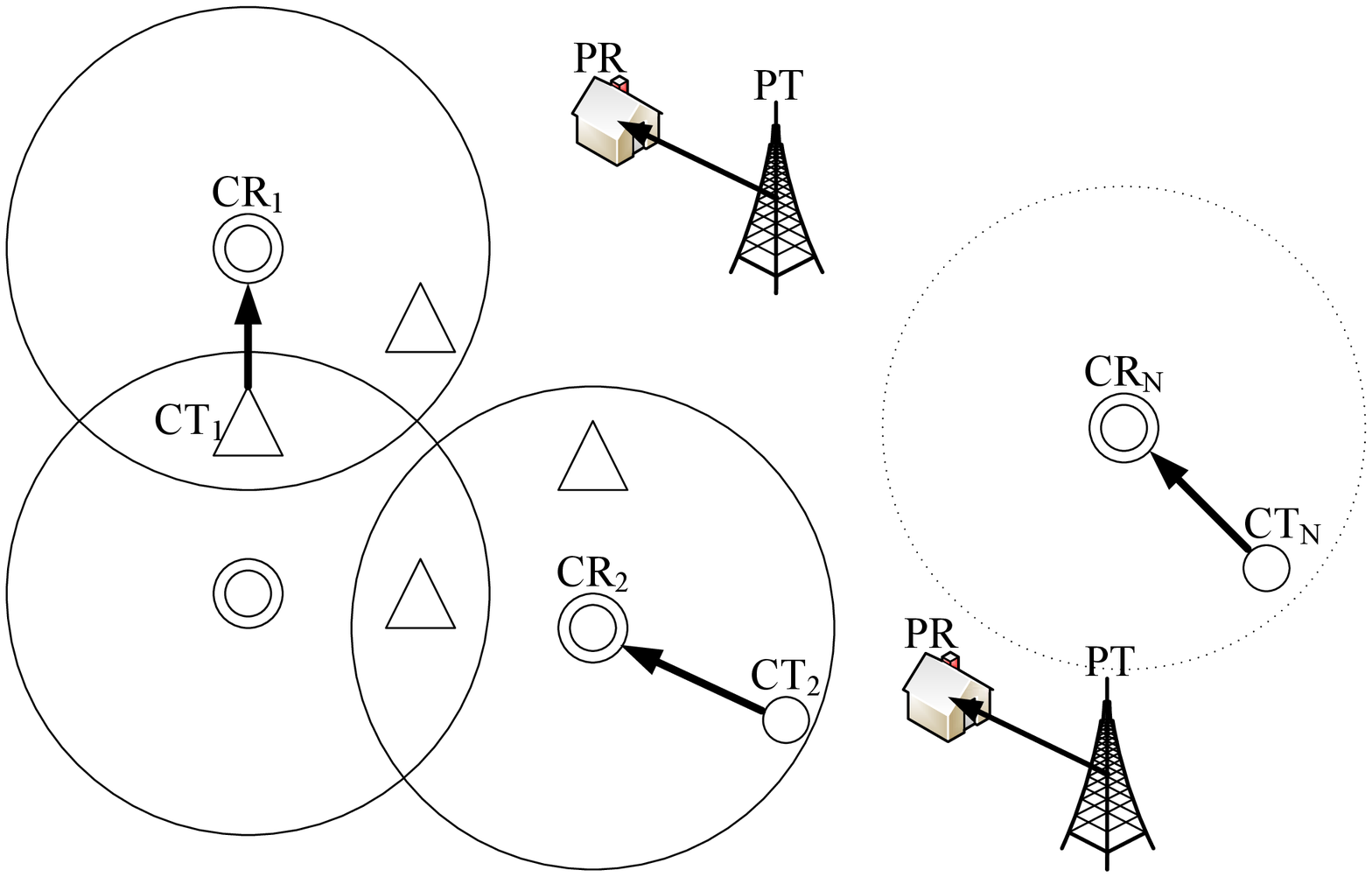}
  \caption{Cognitive wireless mesh networking (\emph{CogMesh}) scenarios.}
\end{figure}

\begin{figure}
  \centering
  \includegraphics[width=0.7\textwidth]{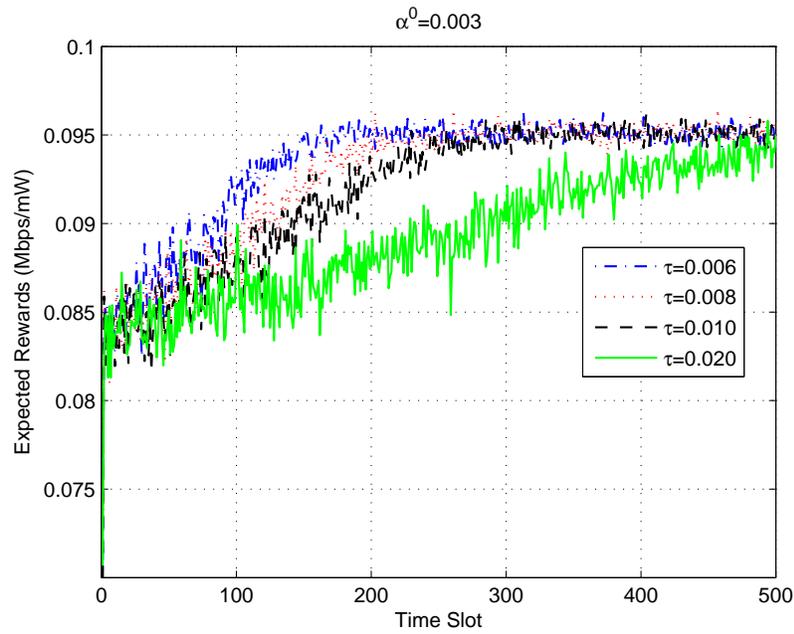}
  \caption{Performance, when $\kappa=0.5$: Impact of the temperature $\tau$ to expected rewards achieved by SU 1.}
\end{figure}

\begin{figure}
  \centering
  \includegraphics[width=0.7\textwidth]{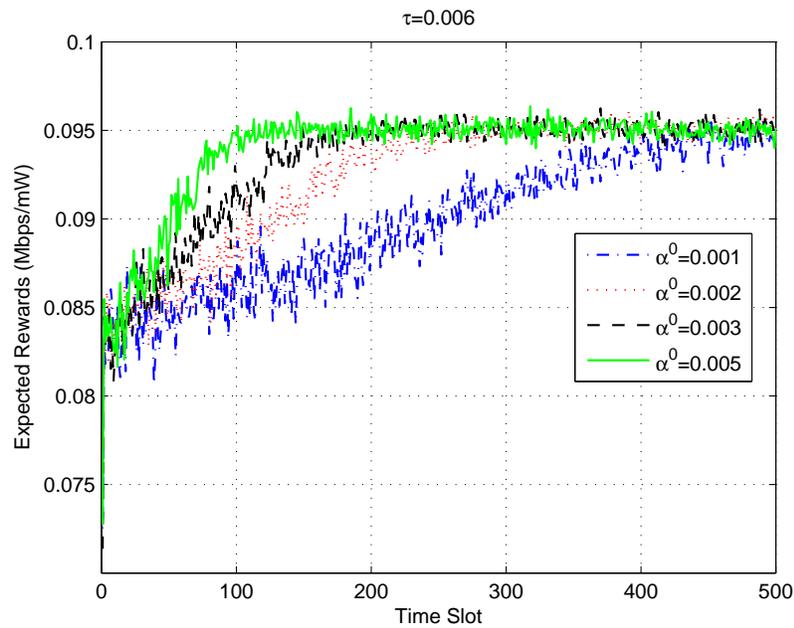}
  \caption{Performance, when $\kappa=0.5$: Impact of the learning rate $\alpha^{0}$ to expected rewards achieved by SU 1.}
\end{figure}

\begin{figure}
  \centering
  \includegraphics[width=0.7\textwidth]{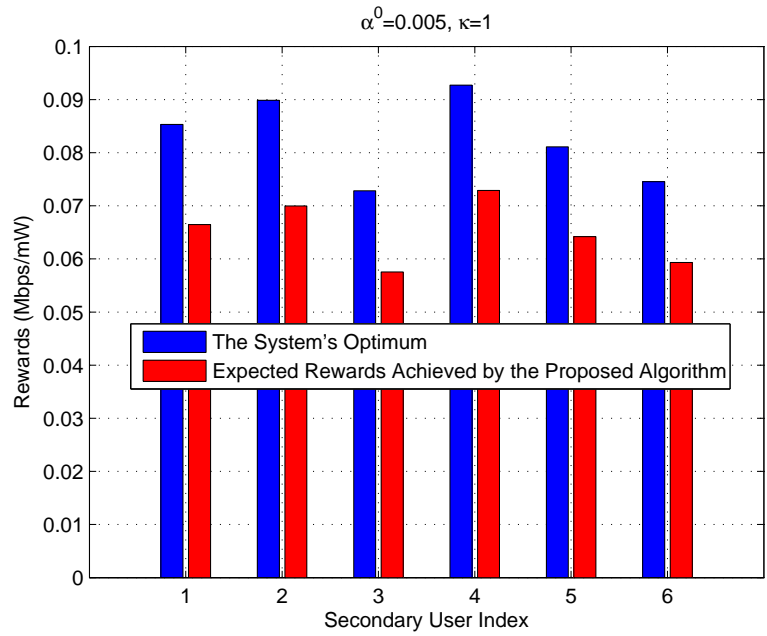}
  \caption{Performance comparison between the proposed algorithm and the system's optimum.}
\end{figure}

\begin{figure}
  \centering
  \includegraphics[width=0.7\textwidth]{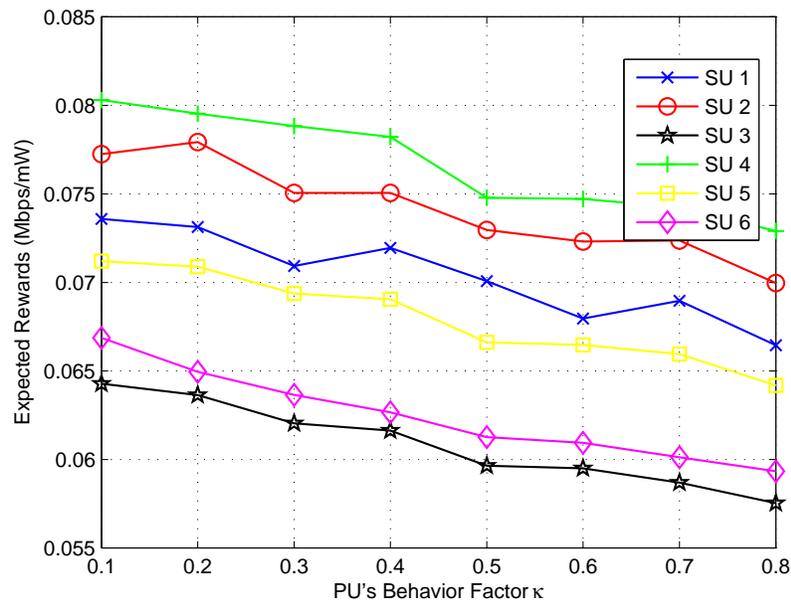}
  \caption{The expected rewards of the SUs versus the PU's behavior factor $\kappa$.}
\end{figure}

\end{document}